\documentclass[arxiv]{paper}

\begin{document}

\ifarxiv
    \title{Protected Data Plane OS Using Memory Protection Keys \\ and Lightweight Activation}
\else
    \title{\sys: High Performance OS I/O Architecture with Protected Data Plane and Lightweight Activation}
\fi

\ifarxiv
    \author[1]{Yihan Yang}
    \author[3]{Zhuobin Huang}
    \author[2]{Antoine Kaufmann}
    \author[1]{Jialin Li}
    \affil[1]{National University of Singapore}
    \affil[2]{Max Planck Institute for Software Systems (MPI-SWS)}
    \affil[3]{University of Electronic Science and Technology of China}
\else
    \author{\small Anonymous Submission \#803}
\fi

\maketitle

\begin{abstract}
    Increasing data center network speed coupled with application requirements
    for high throughput and low latencies have raised the efficiency bar for
    network stacks.
    To reduce substantial kernel overhead in network processing, recent
    proposals bypass the kernel or implement the stack as user space OS service
    – both with performance isolation, security, and resource efficiency
    trade-offs.
    We present \sys, a new network stack architecture that combines the
    performance and resource efficiency benefits of kernel-bypass and the
    security and performance enforcement of in-kernel stacks.
    \sys runs the OS I/O stack in user-level threads that share both address
    spaces and kernel threads with applications, avoiding almost all kernel
    context switch and cross-core communication overheads.
    To provide sufficient protection, \sys leverages x86 protection keys (MPK)
    extension to isolate the I/O stack from application code.
    And to enforce timely scheduling of network processing and fine-grained
    performance isolation, \sys implements lightweight scheduler activations
    with preemption timers.
\end{abstract}

\section{Introduction}
\label{sec:intro}

Modern data center applications and hardware have posed unprecedented challenges
to operating system I/O subsystems.
User requests for these applications are commonly served by hundreds, or even
thousands of services~\cite{fbmemcache, kwiken}.
To ensure responsiveness, each service needs to provide both \textit{low} and
\textit{predictable latency}~\cite{tlatency}, with tail latency service-level
objectives (SLO) commonly in the range of tens to hundreds of
microseconds~\cite{lowlatency}.
At the same time, operators co-locate multiple services on the same physical
host to fully utilize precious computing resources, including CPU cycles,
memory, and network bandwidth.
Consequently, resource efficiency and performance isolation are critical
metrics for data center systems.

Strict latency requirements, coupled with increasing network hardware speed,
have inspired an influx of work seeking to improve OS I/O subsystem performance.
One line of work~\cite{mtcp, sandstorm, arrakis, zygos, demikernel}
implements the I/O stack in user-space libraries that bypass the kernel in the
common case.
In exchange for reduced context switching overhead, these solutions compromise
or completely forgo protection of the I/O stack from untrusted applications.
For better scalability, they use hardware-based packet steering (e.g., receive
side scaling) and run-to-completion on each core, leading to non-preemptive,
multi-queue FIFO scheduling.
Queuing theory dictates that such scheduling policy suffers from higher tail
latency due to head-of-line blocking, particularly for highly skewed workloads
or workloads with high dispersion~\cite{tail-latency, shinjuku}.

In an attempt to address the limitations of kernel-bypass, another line of
work~\cite{snap, tas} implements centralized I/O stack OS services running on
dedicated CPU cores.
The stack is protected and isolated from untrusted applications using process
isolation.
Recent work~\cite{shinjuku, shenango, caladan} goes a step further: They
leverage centralized scheduling services to realize preemptive, single-queue
FIFO, leading to better tail latency.
Unfortunately, an inherent issue for a centralized architecture is degraded
performance isolation, as multiple applications are sharing and contending for
the shared OS service.
By design, these services allocate entire cores for network processing and
scheduling.
Such coarse-grained allocation leads to suboptimal resource utilization, as
the service cores are either a system bottleneck if fully saturated, or waste
CPU cycles when under-utilized.

In this work, we address this dilemma by proposing \sys, a new data plane OS
design.
Our goal is to simultaneously offer strong protection, high throughput, good
tail latency guarantees, efficient utilization of resources, and strict
performance isolation.
Similar to prior kernel-bypass systems, \sys runs the data plane OS in a
distributed fashion: It spawns an instance of the data plane for each
application, and shares the same address space and CPU cores as the application.
By multiplexing CPU cores between the data plane and the application, \sys
utilizes CPU resources more efficiently;
by distributing I/O processing and scheduling, \sys ensures strong
performance isolation between applications.
To protect the data plane from untrusted applications, \sys leverages Intel
MPK~\cite{mpk}, a hardware-based user-space memory protection scheme, to ensure
strong memory isolation.
Our new protection scheme enables the data plane OS to safely run in
user-level threads, with a context switching overhead of only 70 CPU cycles.
To prevent untrusted applications from escalating privileges, we also propose
hardware mechanisms to safeguard MPK registers, without mode switches in the
critical path.

Low and predictable tail latency requires fine-grained preemptive
scheduling~\cite{shinjuku}.
The high overhead of the OS kernel scheduler, coupled with our user-level
threading approach, makes it a challenging task.
In \sys, we address this challenge by implementing lightweight scheduler
activations~\cite{schedactv}.
We use the high resolution timer subsystem in Linux, and implement a \sys
kernel module that efficiently redirects timer interrupts together with the
preempted context, into the user-space data plane scheduler.
\sys also maps local Advanced Programmable Interrupt Controller (APIC)
registers into the data plane OS for protected finer control over preemption
intervals.
Our activation mechanism incurs an overhead of 1098 CPU cycles, allowing
\sys to perform preemptive scheduling every 10{}$\mu$s.

We compare \sys with both the kernel-bypass and central service
architectures on a variety of application workloads.
\sys achieves performance comparable to other performance oriented network
stacks on small-sized RPCs.
On workload with bimodal service times, \sys outperforms IX by 6.9$\times$ in throughput under a tail latency SLO.
When allocated a fixed number of CPU cores, \sys achieves 1.6$\times$ higher
throughput than TAS due to its more efficient CPU utilization.
When co-locating a latency-critical (LC) and a high-throughput (HT) application,
\sys provides better performance isolation than the central service architecture
-- 83.8\% lower tail latency for LC and 16.8\% higher sustainable throughput for HT.

\section{Background and Motivation}
\label{sec:background}

In this section, we outline the goals of a modern data center OS network stack to meet these demands, and discuss deficiencies of existing solutions.

\subsection{Goals for Data Center Network Stacks}
\label{sec:background:goals}

\paragraph{High-Performance.}
Modern data center applications, such as high-performance key value stores, web servers, and in-memory databases, require high message rates.
Small message workloads are common as many applications use remote procedure calls (RPCs) to coordinate distributed components or invoke external services, leading to many small control messages.
Moreover, applications require RPCs to complete in tens to hundreds of $\mu$s, even in the \textit{tail case} and when servers are processing high traffic volumes.

\paragraph{Performance Isolation.}
Data center and cloud systems run concurrent tasks, be it OS and applications,
separate applications, different tasks in an individual application, or requests
on multiple client connections.
Workload characteristics across different tasks are \textit{diverse}:
Some tasks have low throughput but strict latency requirements, while others issue large batches of requests;
some tasks perform simple computation or memory lookups, while others involve long-running jobs such as table scans and garbage collection.
Applications require mechanisms to effectively isolate operations
from performance interference from other ongoing tasks.

\paragraph{Resource Efficiency.}
Beyond performance, the cost of running data center applications is another critical metric for both users and operators.
Cost is directly related to resource efficiency, i.e., how much hardware resources are required to achieve the service-level agreement (SLA).
As such, a crucial goal of the system is to efficiently use resources such as CPU cycles and memory, and to avoid leaving resources idle when there is pending work.

\paragraph{Multi-Tenant Enforcement.}
Finally, a critical responsibility of an OS network stack is \textit{protecting} mutually non-trusting or buggy applications.
The network stack has raw access to network hardware, a shared system resource.%
Without protection, malicious or buggy applications can violate integrity or confidentiality of other applications' network traffic, and interfere with their network performance.
Moreover, correct operation and protocol conformance of the stack requires \textit{timely} handling of protocol events, e.g., acknowledgements, re-transmissions, and congestion control.

\subsection{Prior Network Stack Architectures}
\label{sec:background:prior}

\begin{table}[t]%
  \footnotesize%
  \begin{tabular}{p{1.7cm}cccc}
    \toprule
    \textbf{Network \newline Stack} &
      \multicolumn{1}{p{0.6cm}}{\textbf{High \newline Perf.}} &
      \multicolumn{1}{p{1.1cm}}{\textbf{Perf. \newline Isolation}} &
      \multicolumn{1}{p{1.2cm}}{\textbf{Resource \newline Efficiency}} &
      \multicolumn{1}{p{1.5cm}}{\textbf{Multi-Ten. \newline Enforcement}} \\
    \midrule
    Linux & \xmark & (\cmark) & \cmark & \cmark \\
    IX & (\cmark) & \xmark & \cmark & \cmark \\
    mTCP & \cmark & \xmark & \cmark & \xmark \\
    TAS & \cmark & \xmark & \xmark & \cmark \\
    \textbf{\sys} & \cmark & \cmark & \cmark & \cmark \\
    \bottomrule
  \end{tabular}%
  \caption{Network architectures compared by design goals.}%
  \label{tab:stackarchs}%
\end{table}

A long line of prior work~\cite{arrakis,ix,mtcp,demikernel,zygos,shinjuku,shenango,caladan,tas,snap} explores network stack architectures with different trade-offs among these goals.
We divide these architectures into three major categories: in-kernel, decentralized kernel-bypass, and central services.
\autoref{tab:stackarchs} lists and compares the trade-offs with representative systems for each architecture.

\paragraph{In-Kernel.}
The Linux in-kernel network stack performs poorly on small RPC throughput and latency~\cite{ix,arrakis}.
This is primarily due to expensive context switches between applications and the kernel, and the large code footprint of a general purpose stack.
The millisecond kernel scheduling quantum also leads to long tail latency for short $\mu$s-scale requests.
However, in-kernel network stacks can guarantee protection for multi-tenant enforcement.
The processor's \emph{kernel-mode} prevents applications from directly accessing or manipulating the network stack.
Combined with \emph{preemptive scheduling} with hardware timers, an in-kernel stack also ensures correct and timely execution of protocol events, as well as fair scheduling of different tasks, albeit at a coarse granularity.
By centrally managing all resources, the kernel can also make idle resources available to applications with demand for capacity.

The IX~\cite{ix} protected data plane retains similar protection through CPU virtualization features~\cite{dune}.
IX's custom non-POXSIX API amortizes context switching and protection overhead through \emph{batching}, and improves latency by requiring applications to process incoming requests to completion in order.
However, IX implements \emph{non-preemptive and decentralized scheduling} of applications and the network stack, which increases tail latency and precludes fine-grained performance isolation at connection or request level~\cite{zygos,shinjuku}.

\paragraph{Decentralized Kernel-Bypass.}
Kernel-bypass network stacks, such as mTCP~\cite{mtcp}, Arrakis~\cite{arrakis}, or Demikernel~\cite{demikernel}, go a step further and avoid context switching by implementing the network stack as user space libraries that applications invoke through function calls.
To avoid synchronization, these stacks use hardware packet steering
to ensure packets arrive at the correct application core and then process them locally.
Implemented in a user space library OS, they can also be tailored to and optimized for specific application needs~\cite{exokernel}.

While reducing common-case latency and increasing throughput, kernel-bypass stacks fall short in fine-grained performance isolation due to \textit{cooperative scheduling} that results in first-come-first-serve request processing on each core.
More critically, these network stacks forgo protection from untrusted applications.
Application code can directly manipulate protocol state or delay protocol processing.
Hardware-based solutions~\cite{arrakis} using SR-IOV and IOMMUs only offer limited and coarse-grained protection and isolation, such as filtering packets based on addresses and enforcing rate limits.

\paragraph{Central Service.}
Inspired microkernels, TAS~\cite{tas} and SNAP~\cite{snap} run the network stack as a separate service on \emph{dedicated} processor cores.
Applications interface with the network stack through fast shared memory message passing.
While incurring a small cache coherence overheads for message passing, central service stacks avoid context switching, improve cache locality, and prevent interference from applications.
As a result, they offer throughput and common-case latency comparable to kernel-bypass.

However, shared service architectures complicates performance isolation as they handle network processing for all applications and connections.
Finally, resource efficiency is a major concern;
while services can dynamically add or remove network stack cores, this occurs at coarse granularity, both spatially --- entire cores are allocated --- and temporally --- reallocation occurs at millisecond granularity.
Sharing cores with applications to improve resource utilization, forgoes the performance benefit of this architecture~\cite{snap,tas}.

\subsection{A Dilemma}
\label{sec:background:dilemma}

We now present a series of motivating experiments to demonstrate an unfortunate dilemma:
neither decentralized kernel-bypass nor central service architectures satisfies all our goals outlined in \autoref{sec:background:goals}.
The full experiment setup is detailed in \autoref{sec:eval}.

\begin{figure*}[t]
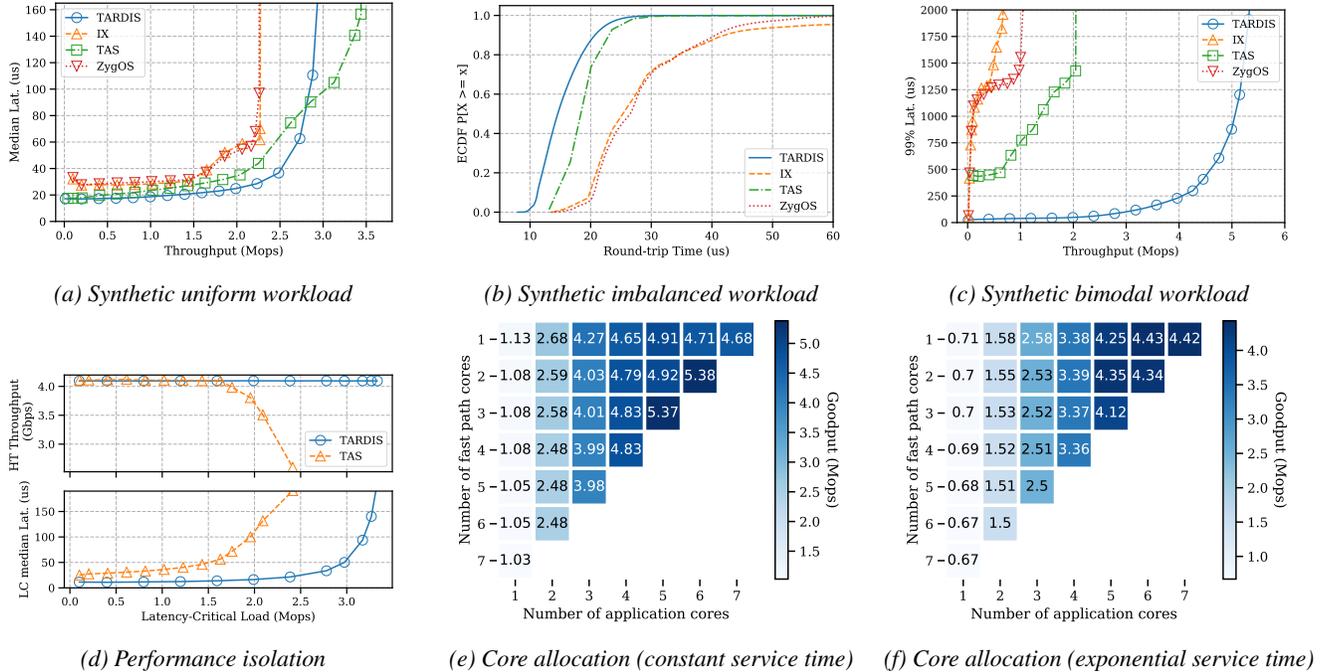

  \centering
  \begin{subfigure}[b]{0.33\textwidth}
      \addgraph{eval-synthetic-uniform}
      \caption{Synthetic uniform workload}
      \label{fig:motiv-uniform}
  \end{subfigure}
  \begin{subfigure}[b]{0.33\textwidth}
      \addgraph{eval-synthetic-imbalanced}
      \caption{Synthetic imbalanced workload}
      \label{fig:motiv-imbalanced}
  \end{subfigure}
  \begin{subfigure}[b]{0.33\textwidth}
      \addgraph{eval-synthetic-bimodal}
      \caption{Synthetic bimodal workload}
      \label{fig:motiv-bimodal}
  \end{subfigure}
  \begin{subfigure}[b]{0.33\textwidth}
    \addgraph{eval-isolation}
    \caption{Performance isolation}
    \label{fig:motiv-isolation}
  \end{subfigure}
  \begin{subfigure}[b]{0.33\textwidth}
    \addgraph{motiv-tas-const}
    \caption{Core allocation (constant service time)}
    \label{fig:motiv-res-const}
  \end{subfigure}
  \begin{subfigure}[b]{0.33\textwidth}
    \addgraph{motiv-tas-exp}
    \caption{Core allocation (exponential service time)}
    \label{fig:motiv-res-exp}
  \end{subfigure}
  \caption{In (a) -- (c), a single application runs on 16 server cores.
  The application performs a synthetic workload with different service time in each experiment: uniform distribution with a mean of 2.5$\mu$s in (a), exponential distribution with 5 cores receiving 10 times more traffic in (b), and bimodal (99.5\%--1, 0.5\%--1000) distribution in (c).
  In (d), two applications collocate on the same server.
  The high-throughput (HT) application sends request at a fixed 4 Gbps rate.
  Load of the latency-critical (LC) application gradually increases.
  In (e) and (f), maximum throughput for different core allocation schemes of a central service stack is shown.
  The application runs a constant service time workload in (e), and an exponential workload in (f).
  }
  \label{fig:motiv}
\end{figure*}

\paragraph{Poor Tail Latency.}
We first show that existing decentralized kernel-bypass stacks suffer from long tail latency when application workloads exhibit high variance in service time or load distribution.
We deploy a server application that performs dummy work on each request, and control the duration of the work to simulate various service-time distribution.
The application runs on 16 CPU cores, and we gradually increase the client load.
We first run a workload with exponential service-time distribution, but unevenly distributes the requests: 5 cores receive 136 Kops while the remaining cores process 13.6 Kops.
For a kernel-bypass stack that implements decentralized first-come-first-serve (d-FCFS) scheduling, e.g., IX, the cores that receive higher traffic experience high CPU load, while the remaining cores stay idle.
The effect is shown in \autoref{fig:motiv-imbalanced}: IX suffers a significantly higher tail latency than TAS, which implements centralized FCFS scheduling.

We next switch to a workload that evenly distributes the request, but with a service time that follows a bimodal distribution with 99.5\% of requests take 1$\mu$s and 0.5\% of requests take 1000$\mu$s.
Existing decentralized stacks lack support for \emph{preemptive scheduling}, which is critical to approximate processor sharing (PS) scheduling.
As a result, long-running tasks in the bimodal distribution block subsequent short tasks waiting in the scheduler queue.
This is demonstrated in \autoref{fig:motiv-bimodal}, where IX incurs a high 99\% tail latency even at moderate load.

\paragraph{Weak Performance Isolation.}
While central service stacks excel at tail latency metrics, they fall short in performance isolation guarantees.
To illustrate this issue, we co-locate two types of applications on the same server:
a latency critical application (LC) that generates small TCP messages and has strict tail latency requirement (representative of a fast in-memory key-value store system like memcached~\cite{memcached}),
and a high-throughput application (HT) that transmits larger messages and expects higher overall throughput for shorter completion time(representative of a data analytics applications like Spark~\cite{spark}).
We fix the load of HT at 4 Gbps and gradually increase the load of LC.
We measure both the tail latency of LC and the sustainable throughput of HT, and show the result in \autoref{fig:motiv-isolation}.

For a centralized architecture such as TAS, the two applications are contending over the shared I/O cores, interfering with each other's performance.
Consequently, once the load of LC exceeds 2.0 Mops, throughput of HT starts to drop compared to when running HT alone.
More critically, tail latency of LC is also negatively impacted, as more LC requests are experiencing longer queuing delay.

\paragraph{Low Resource Efficiency.}
Another metric where central service stacks struggle at is efficient utilization of hardware resources.
Due to their coarse allocation of entire cores to I/O processing, CPU resources can be either left idle when there is not enough I/O workload, or become a bottleneck during bursty traffic.
We demonstrate this phenomenon by running a simple benchmark on TAS, in which a maximum of 8 cores are allocated to both the application and the TAS fast path I/O cores.
The application runs a simple echo server with two types workloads: constant and exponential service-time.
We then sweep the configuration space of core allocation, and show the maximum throughput each configuration achieves in \autoref{fig:motiv-res-const} and \autoref{fig:motiv-res-exp}.
The total throughput varies across configurations by more than 400\% even when all the cores are allocated, demonstrating that it is non-trivial to determine the best configuration for a workload.
The issue becomes even more severe in more dynamic situations where application and I/O workloads fluctuate across time.

\section{A Decentralized Protected Data Plane}
\label{sec:approach}

A decentralized architecture naturally avoids two fundamental limitations of
the centralized service approach:
Instead of processing packets with shared resources, the architecture can
leverage NIC hardware packet steering to process packets on the same CPUs where
the application consumes them.
This not only minimizes cache coherence overheads but also effectively
\emph{eliminates I/O performance interference} among competing applications.
Instead of allocating whole CPU cores to tasks, a decentralized archiecture
can flexibly multiplex inividual cores among application code, the I/O stack,
and other OS functionality. %
The architecture therefore can flexibly and dynamically allocate CPU cycles,
\emph{improving CPU resource utilization}.

We argue that the missing or expensive protection and poor tail latency
(\cref{sec:background:dilemma}) in prior decentralized data
planes~\cite{mtcp, arrakis, ix, demikernel}, are not fundamental, but merely
design and implementation limitations.
To overcome these deficiencies, we need to address the following technical
challenges:

\paragraph{Lightweight protection for data plane OS.}
$\mu$s-scale data center applications issue frequent I/O calls leading to
frequent context switches between application and network stack.
In-kernel network stacks leverage CPU kernel-user isolation for confidentiality
and integrity protection of OS data and privileged CPU features.
Micro-kernels use address space isolation to protect the I/O stack.
Both approaches incur significant hardware overhead for context switches.
Achieving our goals, therefore, requires a lighter-weight solution that
efficiently switches between the I/O stack and the application, but without
\textit{compromising protection}.

\paragraph{Fast preemptive scheduling.}
Correct network protocol processing and low, predictable tail latency require
timely preemption of application-level tasks.
Prior decentralized architectures, however, only implement cooperative
scheduling for two reasons:
1) As I/O stack and the application run in the same thead, the kernel scheduler
cannot switch execution between them.%
2) Current preemption mechanisms incur high overhead, and occur at
millisecond-scale intervals.
Consequently, we need to implement efficient decentralized $\mu$s-scale
preemption, while allowing flexible scheduling between I/O and user tasks.

\subsection{Our Approach}
We propose to address these challenges in \sys using a combination of user-level
threading, hardware-assisted memory protection for I/O stack isolation, and
fine-grained timer interrupt coupled with lightweight scheduler activation.

\paragraph{User threading with virtual memory for confidentiality.}
To minimize context switching overhead, \sys runs the I/O stack in
user-level threads that share the same CPU privilege level and address space as
the application.
We ensure \emph{confidentiality} between mutually non-trusting applications
through virtual memory; each application data plane only has access to buffers
and protocol state for that application.
Application visibility into protocol state for its connections is innocuous.

\paragraph{Memory-key integrity protection.}
We augment this and separately protect network stack memory \emph{integrity}
(\cref{sec:design:protection}).
\sys leverages hardware-assisted user-space memory protection through Intel
MPK~\cite{mpk}.
MPK enforces similar page-level protection as conventional virtual memory, while
adding ISA support to efficiently switch protection domains from user-space.
\sys reserves a hardware protection key for data plane OS memory pages.
We propose a new hardware mechanism to safeguard the protection key from
unauthorized application code and through a call-gate-like mechanism enable
controlled transitions.
Compared to privilege mode and address space switches, our MPK-based protection
scheme incurs a fraction of the overhead but only protects integrity.

\paragraph{Preemptive scheduling with scheduler activations.}
Scheduler activation~\cite{schedactv} is a classic solution for kernel
preemption of user-level threads.
However, the original design creates additional activations -- equivalent to
new kernel threads -- on each preemption.
The cost of spawning kernel threads is prohibitively high 152057 at
our required preemption frequency.
\sys instead implements lightweight activations that enable
preemptive scheduling of user-level threads at \textit{microsecond
granularity} (\cref{sec:design:activation}).
\sys leverages high-resolution timers on modern CPUs to trigger $\mu$s-scale
timer interrupts.
A timer callback function, implemented in a kernel module, redirects these
interrupts directly to the user-space data plane scheduler.
To reduce activation overheads, we reuse the current activation and bypass the
kernel scheduler.
\sys also maps local Advanced Programmable Interrupt Controller (APIC) timer
registers directly into the data plane OS, allowing the user-level scheduler to
dynamically configure preemption intervals based on runtime information.
Leveraging lightweight preemption, \sys effectively approximates PS scheduling
and enforces timely I/O stack execution.

\section{\sys Design}
\label{sec:design}

In this section, we present the concrete design of the \sys OS, including
details of data plane protection, fine-grained preemption, lightweight
activation, and scheduling policies.

\subsection{Design Overview}

\begin{figure}
    \addfig{1.0}{design}
    \caption{\sys system architecture}
    \label{fig:design}
\end{figure}

\autoref{fig:design} shows the high-level architecture of \sys.
Similar to prior designs, \sys separates the control plane from the data
plane OS.
The control plane runs in the kernel.
It manages system-wide resources, allocates CPU cores, and grants I/O queues to
the data plane.
The data plane OS contains both an I/O stack and a scheduler, and runs in the
user address space of each application.
\sys spawns \textit{user-level threads} for both the I/O stack and the
application (by overriding \code{pthread\_create()} using \code{LD\_PRELOAD}),
multiplexed over a kernel thread running on each allocated core.
Using user-level threads reduces both context switching and thread creation
overhead.
The latter is particularly important for designs that spawn a thread for
each user request~\cite{arachne, shinjuku, shenango}.
\sys leverages Intel MPK to provide memory isolation for the data plane OS.

The control plane configures NIC flow steering rules to route network packets to
the appropriate RX queues assigned to the data plane \Circled{1}.
Once the I/O thread is scheduled, it polls a batch of packets from the RX queue
\Circled{2}, and processes them through the stack \Circled{3}.
Since the data plane OS is deployed in a per-application manner, the I/O stack
can be optimized and tailored to the application's need~\cite{exokernel, mtcp,
arrakis, sandstorm}.
After the entire batch is processed, the I/O thread yields to the data plane
user-level scheduler \Circled{4}.
The scheduler maintains both a per-core and a global task queue.
A task scheduling policy (\cref{sec:design:sched-policy}) determines how the
scheduler enqueues/dequeues tasks to/from the two queues.
After the scheduler picks an application task to run, it performs a context
switch and downgrades the user-space protection level
(\cref{sec:design:protection}) before switching to the thread \Circled{5}.
This ensures that the application has no access right to memory pages of the
data plane OS.
The application may perform networking I/O during the task.
\sys exposes a data plane system call interface for requesting services in
the data plane OS.
The system call routine upgrades the user-space protection level, calls the
requested service handler \Circled{6}, and downgrades the protection level
before returning to the application \Circled{7}.
When the application thread yields or is blocked, control is returned to the
user-space scheduler, who can choose a different task or the I/O thread to
run.

\sys uses fine-grained preemptive scheduling, with a configurable preemption interval
down to 10$\mu$s.
When timer interrupt occurs \Circled{a}, the kernel interrupt handler performs a
\textit{lightweight scheduler activation} that upcalls into the data plane
user-level scheduler \Circled{b}, passing along the hardware-saved context of
the preempted thread.
Departing from the original scheduler activation design, \sys reuses the
same kernel thread for activation, saving the expensive cost of creating a new
activation in the kernel.
If the user-level scheduler decides that the preempted task has run for too long
or the I/O stack needs to be scheduled (\cref{sec:design:sched-policy}), it
chooses another application thread or the I/O thread \Circled{c} to
run, moving the preempted task to the end of the run queue \Circled{d}.
Otherwise, it resumes execution of the preempted thread \Circled{e}.

\subsection{Data Plane OS Protection}
\label{sec:design:protection}

\sys runs the data plane OS and application threads in the same user address
space.
To protect the data plane, \sys leverages user-space memory protection keys
(MPK) in recent Intel CPUs~\cite{mpk} to implement efficient memory isolation.

\paragraph{MPK primer.}
MPK enables associating groups of memory pages in an address space to one of 16
hardware protection keys.
To control accesses to each page group, MPK introduces a new \reg{PKRU} CPU
register that contains the permission bits for each of 16 protection keys in
the local hardware thread.
To modify group access permissions, a user space program invokes the new
non-privileged \instr{WRPKRU} instruction to update the corresponding bits in
the \reg{PKRU} register.
\instr{WRPKRU} requires no context switches or TLB flushes, and takes less than
20 CPU cycles~\cite{libmpk}.
Moreover, MPK enables a finer, per-thread view of memory access permissions.

\begin{figure}
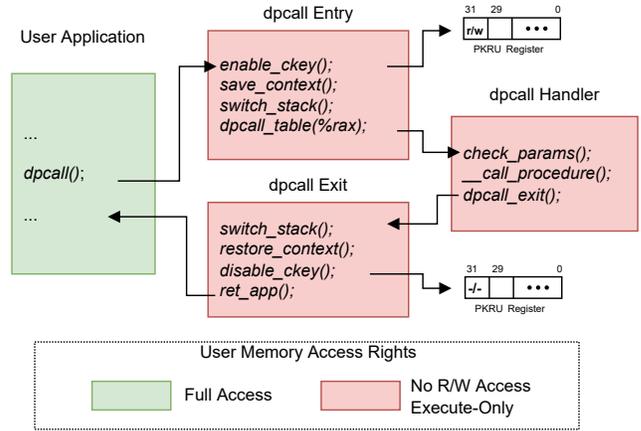

    \addfig{1.0}{dpcall}
    \caption{\sys data plane call procedure.}
    \label{fig:dpcall}
\end{figure}

\paragraph{MPK-based data plane OS protection.}
\autoref{fig:dpcall} shows \sys{}' data plane protection mechanism.
We reserve the last memory protection key, \pkey, for protecting the data plane
OS (bits 30-31 of \reg{PKRU}).
The remaining protection keys are still available for the applications.
When allocating memory to the data plane OS (including code, stack, and data
pages), \sys associates \pkey with each memory page using
\code{pkey\_mprotect()}.
To isolate the data plane from applications, \sys sets the Access Disable
(\textit{AD}) and Write Disable (\textit{WD}) bit for \pkey in the \reg{PKRU}
register when running application threads (\code{disable\_ckey()}).
The MMU hardware ensures that memory regions of the data plane OS are neither
readable nor writable, but still executable, for application threads.
The \textit{AD} and \textit{WD} bits are cleared when entering the data plane OS
(\code{enable\_ckey()}).
To prevent users from circumventing our protection mechanism, we instrument the
\code{mprotect()} and \code{pkey\_mprotect()} system calls to reject attempts to
modify access rights and protection keys for pages in the data plane OS.

When an application thread requests services from the data plane (e.g., TCP
socket \code{send()}) or performs scheduling related operations (e.g., thread
synchronization and yield),
\sys performs a fast user-level context switch to the data plane OS.
We name these fast path system calls \textit{\dpcall{}s}.
\autoref{fig:dpcall} shows the procedure of performing a \dpcall:
The application thread writes the call number and function arguments into
registers, and jumps to a \textit{data plane entry function}.
The entry function is also mapped in the data plane memory region -- application
code can execute the entry point, but are prohibited to read or modify it.
The function clears \textit{AD} and \textit{WD} for \pkey in the \reg{PKRU}
register.
It then saves the application thread context, switches to the data plane stack,
and jumps to the appropriate \dpcall handler using the call number
and an indirection table.
Once the handler finishes serving the request, it calls a \textit{data plane
exit function}, which restores the application thread context, sets \textit{AD}
and \textit{WD} bits for \pkey in \reg{PKRU}, and jumps to the application
thread return address.

\paragraph{Protecting data plane's memory protection key.}
\reg{PKRU} is an unprotected x86 register which can be modified by application
code using non-privileged instruction.
By changing the access bits in \reg{PKRU}, user applications can gain
unwarranted read and write access to the protected data plane memory.
One solution is to apply ERIM~\cite{erim} style binary inspection to prevent
unsafe application code from modifying the \reg{PKRU} register.
Such binary inspection can be integrated into compilers.

\begin{lstlisting}[float, label=code:wrpkru,
caption={\sys \code{WRPKRU} instruction operation. Compared to the original
x86 \code{WRPKRU} instruction, we add instruction pointer tests to restrict
updating \pkey (\reg{PKRU[31:30]}) at the program locations set by the control
plane.}]
IF (ECX = 0 AND EDX = 0)
    IF (RIP = CR9 OR RIP = CR10)
        THEN PKRU := EAX;
        ELSE PKRU[29:0] := EAX[29:0]
    ELSE #GP(0);
FI;
\end{lstlisting}

However, to avoid the expensive binary inspection and including compilers in the
trusted computing base, we propose a lightweight hardware mechanism to prevent
unauthorized updates to \pkey.
We use two 64 bit reserved control registers (e.g., \reg{CR9} and \reg{CR10} on
x86-64) to hold the addresses of the two instructions that update the access
rights for \pkey, one in the data plane entry function and one in the exit
function.
Like other control registers, only the control plane OS running in
privileged-mode can modify their content.
We then modify the \code{WRPKRU} instruction, as shown in \autoref{code:wrpkru},
to perform the following check:
if the program counter does not equal to either of the control registers, the
instruction only updates the lower 30 bits of the \reg{PKRU} register.
The effect of this hardware mechanism is that only the data plane entry and exit
functions can modify the hardware thread's access rights to \pkey.

\paragraph{Security analysis of the \sys data plane.}

Our hardware-based protection mechanism is secure against indirect jump and
return-oriented programming (ROP) attacks.
Only a single instruction in \code{enable\_ckey()} is permitted to modify \pkey
in \reg{PKRU}.
We manually program \code{enable\_ckey()}, \code{save\_context()}, and
\code{switch\_stack()} in assembly code, and ensure only direct jumps are used.
This enforces a \textit{single control flow} between \pkey is updated and the
data plane stack is switched.
Since the data plane stack resides in protected memory, user programs are
prohibited to write return addresses into the stack.
The same mechanism is applied to the \code{switch\_stack()},
\code{restore\_context()}, and \code{disable\_ckey()} sequence.
Consequently, user programs are incapable of escalating their privilege by
running with \textit{AD} and \textit{WD} of \pkey disabled.

Our mechanism, however, suffers the same security vulnerabilities of Intel MPK
itself.
Most notably, Spectre~\cite{spectre} and Meltdown~\cite{meltdown} attacks can
bypass hardware protection of MPK~\cite{transient_exec_attacks}.
\sys therefore offers weaker security guarantees than Linux kernel with
kernel page-table isolation (KPTI)~\cite{kaiser}.

\paragraph{Hardware Implementation and Resource Overhead}

To validate the feasibility of our hardware proposal, we implement our MPK design in RTL in the open-source CVA6~\cite{cva6} 64-bit RISC-V CPU.
CVA6 is a 6-stage, single-issue, in-order processor core.
It implements RISC-V I, M, A, and C extensions, plus three privilege levels M, S, and U.
Unfortunately, current RISC-V extensions have no support for user-level memory protections.
We therefore use the Donky~\cite{donky} modified CVA6 core as the base design, which implements MPK-like memory protection in CVA6.

Our implementation adds two RISC-V supervisor mode CSRs, \code{CSR\_SMPK\_ENTRY} and \code{CSR\_SMPK\_EXIT}, which content can only be modified in supervisor mode.
We reserve the last protection key slot (slot 3, bits [43:33]) in the MPK register (\code{CSR\_MPK}) to \sys.
When updating the MPK register in user mode, we additionally check the program counter (\code{pc\_i}) against the two added CSRs.
If the program counter differs from the content of both registers, the reserved protection key remains unmodified.

\begin{table}
    \centering
    \begin{tabular}{p{14mm}|cccc}
        CPU & LUT & Register & BRAM & DSP \\ \hline
        Donky & 30.97\% & 11.73\% & 11.24\% & 3.21\% \\
        \sys & 31.07\% & 11.76\% & 11.24\% & 3.21\% \\ \hline
        Overhead & 0.10\%\small{(205)} & 0.03\%\small{(128)} & 0.00\% & 0.00\% \\
    \end{tabular}
    \caption{FPGA resource usage of \sys RISC-V CPU core}
    \label{tab:hw-res}
\end{table}

We synthesized our hardware design using Xilinx Vivado~\cite{vivado} $2021.2$.
The hardware target is a Digilent Genesys2~\cite{genesys2} board which equips with a Xilinx Kintex-7 XC7K325T FPGA.
\autoref{tab:hw-res} shows the generated resource utilization report.
Compared to the original Donky design, \sys only uses $0.10\% (205)$ and $0.03\% (128)$ additional LUT and register resources on the FPGA.
Moreover, our hardware modifications do not slow down the processor: the frequency of all the generated clocks remains unchanged from the original design.
The synthesis result testifies that our MPK proposal is highly feasible on real hardware.
This should not come as a surprise, given the simplicity of the design.

\subsection{Preemptive Scheduling and Lightweight Activation}
\label{sec:design:activation}

\sys relies on frequent preemption of user-level threads to ensure strong
tail latency guarantee and timely execution of the I/O stack.
Existing timer interrupt and preemptive scheduling mechanisms in Linux, however,
are inadequate for two reasons:
First, the high overhead of running the kernel scheduler and context switching
among kernel threads limits the preemption interval to milliseconds, much longer
than our $\mu$s-scale tail latency target.
Second, the kernel scheduler has no visibility of user-level threads.
After a preemption, the scheduler is incapable of switching to other ready user-level
threads in the same kernel thread.

Scheduler activation~\cite{schedactv} is a classic solution to the system
integration issues of user-level threads.
However, the original design creates additional activations -- equivalent to
new kernel threads -- on each preemption.
The cost of spawning kernel threads is prohibited high 152057 at our
intended preemption frequency.
To address this challenge, we implement a \textit{lightweight activation}
mechanism to enable fine-grained preemptive scheduling of user-level threads.
Since our target preemption interval is much lower than the kernel jiffy, we
leverage the high resolution timer (\texttt{hrtimer}) system provided by
Linux~\cite{hrtimer}, which internally uses the Advanced Programmable Interrupt
Controller (APIC) timers on x86 machines.

\paragraph{Lightweight activation mechanism.}
The \sys control plane registers a \texttt{hrtimer} callback function with
the OS kernel when initializing the data plan OS.
\sys uses this callback function as a \textit{trampoline} to jump to the
data plane user-level scheduler.
When the \sys callback is invoked by the kernel \texttt{hrtimer} IRQ handler,
it stores the context of the interrupted thread (saved by the hardware interrupt
handler) into the stack of the data plane OS.
It then modifies the saved context to that of the data plane OS, with the return
address set to the \texttt{entry} function of the data plane scheduler.
Once the kernel completes the interrupt and returns to user-space, it restores
the modified context and jumps to the data plane scheduler, instead of the
original interrupted user-level thread.
The data plane scheduler finds the context of the interrupted user-level thread
in its stack, and decides if it should switch to a different user-level thread
to run (\cref{sec:design:sched-policy}).
The I/O stack can additional register timers (e.g., for TCP retransmission) with
the data plane scheduler.
The scheduler switches to the I/O thread when one of the timers expire.

\begin{table}
    \centering
    \begin{tabular}{p{14mm}|cccc}
        & POSIX signal & IPI & \sys \texttt{hrtimer} \\ \hline
        Sender & 2418 & 42077.95 & - \\
        Receiver & - & 884.67 & 1097.92 \\ \hline
        Total & 5645 & 32417.18 & 1097.92 \\
    \end{tabular}
    \caption{Interrupt overhead in CPU cycles for POSIX signal, IPI, and \sys \texttt{hrtimer}-based activation.}
    \label{tab:interrupt}
\end{table}

Our hrtimer-based preemption scheme is faster than alternative approaches.
For instance, the centralized dispatcher in Shinjuku~\cite{shinjuku} uses POSIX
signals or inter-processor interrupts (IPIs) to preempt worker threads.
The overhead of processing signals or IPIs in Linux, however, is substantial.
In \autoref{tab:interrupt}, we compare the performance (in CPU cycles) of the
various approaches to deliver interrupts to the data plane scheduler.
We show in \cref{sec:eval:micro} that the low-overhead of our activation
mechanism enables \sys to perform preemptive scheduling at 10$\mu$s
granularity.

\paragraph{Configurable preemption in the data plane OS.}
The above design works well for preemption that occurs at regular interval:
The control plane configures the interval in the callback function, and when the
callback is invoked, it resets the \texttt{hrtimer} to expire after the
interval.
A fixed interval, however, may lead to many unnecessary interrupts that
negatively impact the efficiency of the system.
As we later explain in \cref{sec:design:sched-policy}, \sys only preempts a
task if its runtime exceeds a threshold.
For shorter tasks, handling highly frequent interrupts simply waste CPU cycles.
We therefore augment the data plane scheduler to dynamically configure hrtimer
preemption.
Specifically, the control plane maps local APIC timer registers
(\texttt{APIC\_TDCR}, \texttt{APIC\_TMICT}, and \texttt{APIC\_TMCCT}) into data
plane OS memory during initialization.
Writing to these registers allows the data plane to directly set, reset, and
cancel APIC timers.
When scheduling a task to run, the scheduler sets the hardware timer to expire
after a threshold; it cancels the timer if the task completes before the
threshold, avoiding unnecessary interrupts.
Since the registers are mapped to data plane OS memory, our protection mechanism
(\cref{sec:design:protection}) ensures hardware timers are inaccessible to
user programs.
Note that in the more recent x2APIC~\cite{x2apic} architecture, APIC registers
can only be accessed through Model Specific Register (MSR) interfaces, which
require privileged instructions.
As a workaround, we explicitly disable x2APIC in the kernel parameters
(\code{nox2apic}).

\subsection{Task Scheduling Policy}
\label{sec:design:sched-policy}

One benefit of running the data plane scheduler in user-space is that the
scheduling policy can be tailored to the requirements of the target
application~\cite{ghost}.
In this work, we implement a scheduling policy that offers good tail latency
guarantees across a wide range of application workloads.

\paragraph{Approximating c-FCFS and PS scheduling.}
As we argued in \cref{sec:background}, a scheduling policy that combines c-FCFS
and PS theoretically gives the best tail latency behavior for many classes of
service time distribution.
To approximate c-FCFS, the data plane scheduler maintains both a core local- and
a global task queue.
Note that the global queue is only for threads within the application process.
When new user-level threads are created or become ready, they are enqueued to
the tail of the global task queue.
The data plane scheduler runs threads from the local queue in FCFS order, and if
the local queue is empty, it pops up to $N$ threads from the global queue and adds
them to the local task queue.
$N$ is a configurable parameter that controls the trade-off between scheduler
efficiency and accuracy of c-FCFS approximation:
When a smaller $N$ is used, tasks are more likely to be scheduled in the order
they arrive, but the scheduler suffers higher synchronization overhead
contending on the global task queue;
when using a larger $N$, frequency of thread synchronization is reduced, but the
task scheduling deviates more from c-FCFS towards a d-FCFS policy.

To approximate PS scheduling, the data plane OS defines a time quanta $T$.
When the data plane scheduler receives a timer activation
(\cref{sec:design:activation}), it checks if the current running thread has been
scheduled for more than $T$.
If the thread has exceeded the quanta, the scheduler pops the task from the
local queue, enqueues it to the end of the global task queue, and schedules the
thread at the head of the local queue.

\paragraph{Choosing between I/O and application threads.}
The data plane scheduler has two options when making scheduling decisions:
picking an application thread to run, or scheduling the data plane I/O thread.
To ensure good application-level tail latency behavior, application threads
typically are given higher scheduling priority than the I/O thread.
However, there are two cases where the I/O thread takes precedence.
Firstly, certain I/O tasks are time sensitive and need to be scheduled with
real-time guarantees.
Examples include TCP acknowledgement and retransmission.
For these tasks, the I/O stack registers special timers with the scheduler.
The scheduler runs the I/O thread if any of the timers has expired, even when
application threads are ready.
Second, long-running application tasks, such as garbage collection, should not
block I/O processing indefinitely.
This ensures both high I/O performance and \textit{predictable} latency for
short tasks.
Towards this goal, we apply a dynamic priority adaptation policy.
When an application task is preempted and placed at the end of the run queue
(i.e., they have exceeded their scheduling quota), the task is marked as a low
priority job.
When the head of the queue is a low priority task, the scheduler runs the I/O
thread, and only schedules the task after the I/O thread completes.

\section{Evaluation}
\label{sec:eval}

We implemented \sys in 12863 lines of C and 150 lines of assembly code,
including both the control plane and the data plane OS.
The \sys kernel module is written in 301 lines of C code.
Our testbed consists of two servers each equipped with dual-socket 16-core Intel
Xeon Gold 6326 2.90 GHz CPUs, 256 GB RAM, and a Mellanox ConnectX-5 EN 100 GbE
single-port NIC.
The two servers are directly connected by a 100 GbE QSFP28 cable.
For software environment, our servers run Ubuntu Linux 20.04 with kernel version
5.4.0 and DPDK 20.11.3 LTS.
We reserve 2048 2 MB huge pages for DPDK.
When evaluating Linux performance, we pin benchmark processes to CPU cores and
set their scheduler priority to real-time.
Note that when evaluating IX, we use Intel XL710 40Gbps NICs, Ubuntu Linux
16.04 with kernel version 4.4.0, and DPDK version 16.04, as those are the only
configurations supported by IX.

\paragraph{Comparison systems.}
We compare \sys to five other systems: a baseline Linux-based system,
mTCP~\cite{mtcp}, IX~\cite{ix}, ZygOS~\cite{zygos}, and TAS~\cite{tas}.
The Linux-based system runs the in-kernel network stack, which we configured and tuned using
the instructions in \cite{tail-latency} (denoted as \textit{Linux} in the
graphs).
mTCP and IX are recent kernel-bypassed, libOS-style I/O stacks that optimize
host networking performance for data center applications.
To protect the stack from untrusted applications, IX uses hardware
virtualization techniques in Dune~\cite{dune} to run the data plane in VMX
non-root mode.
ZygOS builds upon IX and implements work stealing for better tail latency.
TAS implements a centralized TCP service running in dedicated processes and CPU
cores.
TAS additionally splits TCP processing into a fast-path and a slow-path.

\subsection{Micro-benchmarks}
\label{sec:eval:micro}

To evaluate the basic performance characteristics of \sys, we run a simple
TCP echo benchmark.
We deploy open-loop clients which generate TCP request messages with a fixed
payload size (64 bytes if not otherwise specified).
We run a server process on the other machine that replies each request with the
same payload.
The application runs a single server thread on each server core.
For TAS, we allocate one extra core to run the centralized fast path TCP stack.

\begin{figure*}[ht]
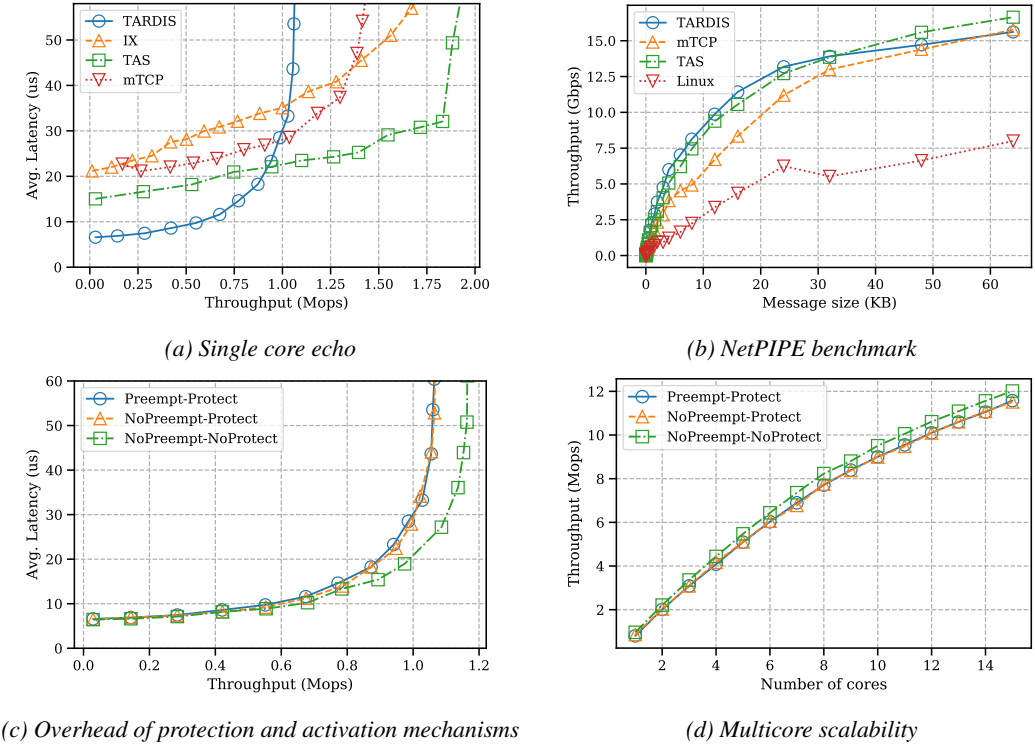

    \centering
    \begin{subfigure}[t]{0.4\textwidth}
        \addgraph{eval-echo}
        \caption{Single core echo}
        \label{fig:eval-echo}
    \end{subfigure}
    \begin{subfigure}[t]{0.4\textwidth}
        \addgraph{eval-netpipe}
        \caption{NetPIPE benchmark}
        \label{fig:eval-netpipe}
    \end{subfigure}
    \begin{subfigure}[t]{0.4\textwidth}
        \addgraph{eval-overhead}
        \caption{Overhead of protection and activation mechanisms}
        \label{fig:eval-overhead}
    \end{subfigure}
    \begin{subfigure}[t]{0.4\textwidth}
        \addgraph{eval-multicore}
        \caption{Multicore scalability}
        \label{fig:eval-scale}
    \end{subfigure}
    \caption{Micro-benchmarks to show performance characteristics of
    \sys. (a) shows tail latency vs.\ throughput when running an open-loop
    TCP echo benchmark. We run a NetPIPE benchmark in (b) to evaluate the
    throughput of \sys with increasing message size. In (c), we disable
    MPK-based protection and scheduler activation in turn to measure their
    impact on \sys performance. In (d), we run the TCP echo benchmark with
    increasing number of CPU cores to measure multicore scalability.}
\end{figure*}

\paragraph{Single core performance.}
We first deploy a single server core, gradually increase the offered client
load, and measure both the average latency and the throughput of each system.
\sys outperforms all comparison systems in latency by a wide margin, even though those systems are highly optimized for single core small-sized I/O.
This level of processing efficiency is attributed to the low overhead of
\sys{}' protection and activation mechanisms.
\sys does attain lower single core throughput than the other systems.
Both mTCP and IX use aggressive batching for higher throughput.
\sys, on the other hand, optimizes for latency.
TAS' high throughput is attributed to the additional CPU resource.
When we allocate a single core to TAS, its throughput drops significantly.

We also use the NetPIPE~\cite{netpipe} benchmark with increasing buffer size to
measure how \sys performs when processing larger messages, and show the
result in \autoref{fig:eval-netpipe}.
\sys attains above 15 Gbps throughput when processing 64 KB messages on a
single core.
This throughput is 0.95$\times$ higher than Linux, and within 6.18\% of TAS which uses two cores.

\paragraph{Overhead of data plane protection and activation.}
To understand the overhead of our protection (\cref{sec:design:protection}) and
activation (\cref{sec:design:activation}) mechanisms, we use the same TCP
benchmark, and test \sys with the following configurations:
1) no protection for the data plane and disable timer activation,
2) enable protection but keep timer activation disabled, and
3) full \sys with both protection and timer activation.
\autoref{fig:eval-overhead} shows the average latency with increasing offered load
for each configuration.

Enabling protection and timer activation has minimum impact on the end-to-end latency of \sys (close to 0\% higher latency).
As discussed in \autoref{sec:design}, the MPK-based protection mechanism adds
around 30{}ns overhead to context switching.
These overheads are dwarfed by the network and other software components in the
end-to-end latency.
For a simple echo application that has near-zero application-level processing,
these overheads have bigger impact on the maximum throughput.
The configuration with both protection and preemption disabled achieves a throughput of 1.16 Mops, translating to 858.4 ns per request on average.
Enabling protection and adding preemptive scheduling reduce the throughput by 7.9\%.
This throughput impact, however, will decrease with higher application-level work.

\paragraph{Multicore scalability benchmark}
To test how well \sys scales, we deploy the same TCP echo application on an
increasing number of CPU cores and measure the maximum attainable throughput.
As shown in \autoref{fig:eval-scale}, throughput of \sys scales close to
linearly to the number of cores.
With RSS, network packets are steered evenly across all the cores, distributing
the I/O work roughly equally to each core.
Since \sys does not steal I/O work across cores, each I/O stack can perform
packet processing independently, minimizing thread synchronization that limits
scalability.
Work stealing done by the scheduler also has minimum scalability impact, as our
simple echo benchmark can hardly lead to skewed application-level work across
cores.

\subsection{Synthetic Workloads}
\label{sec:eval:synthetic}

Next, we test \sys and the comparison systems with a more realistic set of
application workloads.
We use the same set of motivating experiments in \autoref{sec:background:dilemma}:
one server application running on 16 server cores.
The application performs dummy work to simulate various service-time distribution:
\textbf{(a)} a uniform distribution with a mean of 2.5$\mu$s,
\textbf{(b)} a bimodal distribution with 99.5\% of requests take 1$\mu$s and
0.5\% of requests take 1000$\mu$s, and
\textbf{(c)} an imbalanced workload where 5 cores receive 136 Kops of requests while the remaining cores receive request load of 13.6 Kops.

In \autoref{fig:motiv-uniform}, we use the uniform workload, gradually
increase the offered client load, and measure the median latency of each
system.
All the comparison systems are optimized for this simple workload.
Even though \sys pays the cost of MPK-based protection, work-stealing for approximating c-FCFS, and timer activation for preemptive scheduling, it still outperforms IX and ZygOS both in throughput (31\%) and latency (59\%).
\sys achieves equivalent latency as TAS, while attains lower maximum throughput.

\autoref{fig:motiv-bimodal} shows the same tail latency vs.\ offered
load measurement when running our bimodal service-time workload.
As explained in \autoref{sec:background:dilemma}, the lack of preemptive scheduling support in IX leads to long tail latency behavior.
Work stealing in ZygOS alleviates this issue to a certain extent, as idle cores can steal
short tasks that are being blocked.
However, its effectiveness is rather limited beyond moderate client load.
In contrast, \sys aggressively preempts long-running tasks, allowing
subsequent short tasks to complete without suffering long queuing delay.
\sys maintains tail latency below 1000$\mu$s before hitting load saturation (5 Mops), while IX and ZygOS already show tail latency above 1000$\mu$s at load beyond 0.2 Mops.

Lastly, we run our imbalanced service-time workload and show the result in
\autoref{fig:motiv-imbalanced}.
By performing dynamic work stealing, \sys effectively addresses the load imbalance issue caused by d-FCFS scheduling -- it improves 99\% tail latency by more than 6.4$\times$ comparing to IX.
It also outperforms ZygOS in tail latency by 4.1$\times$ as ZygOS lacks preemptive scheduling support, therefore suffering from head-of-line blocking.

\subsection{Resource Efficiency}
\label{sec:eval:resource}

\begin{figure}
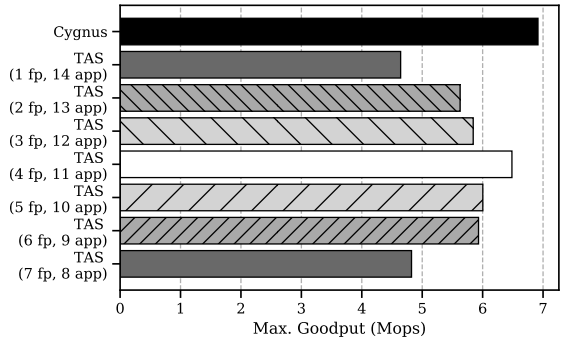

    \addgraph{eval-resource-allocation-15cores}
    \caption{Evaluation of resource efficiency.
    The application simulates an exponentially distributed service-time.
    We measure the maximum throughput achieved by \sys and TAS.
    For TAS, we manually sweep all CPU allocation space.}
    \label{fig:eval-resource}
\end{figure}

To evaluate \sys{}' efficiency at utilizing CPU resources, we run an exponentially distributed service-time workload, and measure the maximum throughput attainable by \sys and TAS.
We allocate 16 CPU cores to both systems.
\sys allocates one core to the control plane, while TAS uses one core for TCP slow path.
\sys allocates the remaining 15 cores to the application.
For TAS, we manually adjust the partition of cores between TCP fast path and the application.
\autoref{fig:eval-resource} shows the maximum throughput attained by each configuration.
For TAS, when less than four cores are allocated to TCP fast path, network processing becomes the bottleneck of the system, while the application cores are under-utilized.
When more than four cores are allocated to TCP fast path, application cores are fully
saturated, but the fast path cores have idle cycles that are not being utilized.
In contrast, \sys is able to efficiently multiplex CPU cores between I/O
processing and application-level work, fully utilizing all the available CPU
resources.
Our scheduling policy and preemption mechanism guarantee that neither I/O nor
application threads will block the other component, ensuring fair sharing of CPU
cycles.
Even comparing to the best configuration of TAS (4 cores for TCP fast path and 11 cores for application), \sys achieves 6.6\% higher throughput.

\subsection{Performance Isolation}
\label{sec:eval:isolation}

Lastly, we evaluate the performance isolation property of \sys, and compare to TAS.
This is the same isolation experiment we used in \autoref{sec:background:dilemma}:
two colocating applications; one latency critical application (LC) that generates small TCP messages, and one high-throughput application (HT) that transmits larger messages and requires higher overall throughput.
We allocate 15 total cores to both \sys and TAS, with one core assigned to the control plane and TCP slow path respectively.
When deploying \sys, we evenly divide the remaining 14 cores to LC and HT.
For TAS, we allocate 2 cores for TCP fast path, and evenly divide the remaining cores to the two applications.
We fix the load of HT at 4 Gbps and gradually increase the load of LC.

As shown in \autoref{fig:motiv-isolation}, the decentralized model of \sys allows I/O processing, scheduling, and application-level work to be physically partitioned between the two applications.
Increase in the LC load therefore does not affect HT at all:
sustainable throughput of HT stays constantly at 40 Gbps, regardless of the LC
load.
At the same time, bursty I/O processing of HT has no impact on the tail latency
of LC.
This is in contrast to the behavior of TAS.
As we explained in \autoref{sec:background:dilemma}, the two applications are contending over the shared fast path cores, resulting in performance interference.
Consequently, \sys achieves a 0.94$\times$ higher sustainable LC load under a tail latency SLO of 100$\mu$s.

\subsection{Memcached}

\begin{figure}
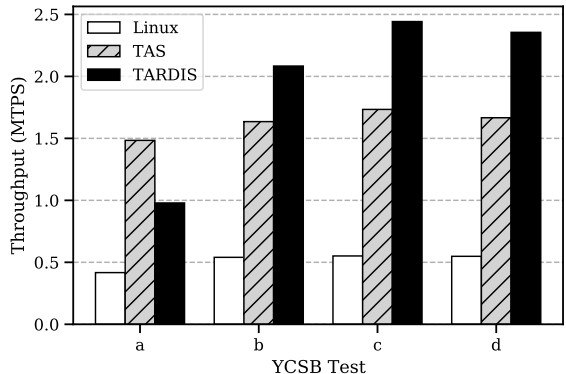

    \addgraph{eval-memcached}
    \caption{Memcached evaluation}
    \label{fig:eval-memcached}
\end{figure}

Lastly, we evaluate how well \sys performs on real world applications.
We use the popular memcached~\cite{memcached} key-value store as our target application, and compare the performance across Linux, TAS, and \sys.
Porting memcached to \sys requires minimum effort, since \sys exposes a POSIX-compliant API.
We only need to instrument the memcached server code to use our threading library.
We allocate four total cores to each system.
For Linux and \sys, we spawn four memcached server threads, each pinned to a dedicated core.
For TAS, we allocate one core for TCP fast path processing and three cores for three memcached threads.
We use the popular YCSB a -- d benchmarks for evaluation.

\autoref{fig:eval-memcached} shows the maximum throughput achieved by each system.
\sys outperforms both Linux and TAS in throughput by a wide margin for three of the four benchmarks.
For instance, in YCSB workload c, a read-only workload, \sys attains 343\% and 41\% higher throughput than Linux and TAS respectively.
The improvement in throughput can be attributed to the fast I/O processing and lightweight protection mechanisms of \sys.
Unfortunately, \sys current has a performance issue in mutex, and thus obtains a lower throughput in YCSB workload a, which contains 50\% write operations.
We are currently working on a fix of the issue.

\section{Related Work}
\label{sec:relwk}

\paragraph{Kernel bypass data plane OS:}
The library operating system~\cite{exokernel, libos} architecture has inspired a
recent line of work that implements data plane OSes in user-space~\cite{arrakis,
ix, zygos, demikernel, mica, chronos}, bypassing the kernel for higher I/O
performance.
Arrakis~\cite{arrakis} leverages hardware I/O virtualization (SR-IOV and IOMMU)
to provide protection to the data plane OS.
However, the approach is constrained by the available virtual functions and
limited coarse-grain resource controls such as rate limiters and
filtering.
IX~\cite{ix} uses CPU hardware virtualization techniques such as
VT-x~\cite{vt-x} to deploy a protected data plane kernel in privileged non-root
ring 0.
ZygOS~\cite{zygos} implements work stealing atop IX to approximate c-FCFS
scheduling for better tail latency.
Demikernel~\cite{demikernel} implements a general-purpose, portable data plane
OS that works on heterogeneous kernel-bypass devices.
Compared to prior kernel-bypass library OSes, \sys offers stronger
protection for the data plane without compromising performance, and enables
aggressive preemptive scheduling for stronger tail latency guarantees.

\paragraph{Centralized I/O and scheduling:}
The non-optimal task scheduling and weaker protection model have lead to
proposals of centralized I/O and scheduling architectures~\cite{tas, snap,
shinjuku, shenango, caladan, ghost}.
TAS~\cite{tas} and Snap~\cite{snap} implement networking I/O as a centralized OS
service, similar to a microkernel architecture, that runs in a separate process
on dedicated cores.
Shinjuku~\cite{shinjuku} uses a centralized core to schedule all
application-level tasks in a c-FCFS manner and preempts long-running tasks to
simulate PS scheduling.
Shenango~\cite{shenango} and Caladan~\cite{caladan} implements a centralized
scheduler that collects control signals and allocates CPU cores to applications
at fine granularity.
As we have demonstrated in \cref{sec:background} and \cref{sec:eval}, a
centralized architecture can lead to non-optimal usage of CPU resources.
\sys addresses this issue by multiplexing CPU cores among I/O, scheduler,
and application work.

\paragraph{I/O stack specialization:}
An orthogonal line of work implements specialized I/O stack in user-space, using
the DPDK~\cite{dpdk} library or RDMA hardware.
Sandstorm~\cite{sandstorm}, mTCP~\cite{mtcp}, eRPC~\cite{erpc},
Andromeda~\cite{andromeda}, and SocksDirect~\cite{socksdirect} fall under this
category.
These work, however, do not offer the protection and manageability benefits of
an operating system.
The I/O stack in \sys can apply optimization techniques proposed by these
work to further improve performance.

\paragraph{User-level threading:}
User-level threading system~\cite{arachne, capriccio, task-manage} offers better
performance and predictability than kernel threads, thus becoming an attractive
option for deploying microsecond-scale applications~\cite{demikernel, shinjuku}.
Prior user-level threading systems, however, only support cooperative
scheduling, which can result in poor tail latency and delaying of time-sensitive
I/O tasks.
\sys addresses this issue by implementing fine-grained preemptive scheduling
for user-level threads using lightweight timer activation.

\section{Conclusion}
\label{sec:conclusion}

In this paper, we described \sys, a new OS networking architecture that
provides high performance but without compromising protection, functionality, or
manageability. Unlike previous approaches, \sys runs the OS networking
service in shadow user-level threads, and protects the network stack through
memory isolation using Intel Memory Protection Keys. We have also implemented
I/O optimized scheduler activations to enforce correct and timely execution of
network protocols in user-level threads.
 
\bibliography{paper}

\end{document}